\documentclass{aa}  

\usepackage[normalem]{ulem}
\usepackage{graphicx}
\usepackage{color}
\usepackage{url}
\usepackage[export]{adjustbox}
\usepackage[colorlinks=true, allcolors=blue]{hyperref}
\usepackage{lipsum}
\usepackage{multirow}
\usepackage{xcolor}
\usepackage{orcidlink}
\usepackage{xspace}
\usepackage{soul}
\usepackage{tablefootnote}
\usepackage[switch]{linenoaa} 
\usepackage[referable]{threeparttablex}

\usepackage{txfonts}
\usepackage{natbib}
\usepackage{amsmath}    
\usepackage{amssymb} 
\usepackage{bm}
\usepackage{mathtools}
\usepackage{orcidlink}
\bibpunct{(}{)}{;}{a}{}{,}

\usepackage{graphicx}

\begin{document}

\title{The constraining power of X-ray polarimetry: detailed structure of the intrabinary bow shock in Cygnus X-3}

\subtitle{}

\titlerunning{Cygnus X-3 intrabinary bow shock}

\author{Varpu Ahlberg\inst{1}\orcidlink{0009-0006-9714-5063}
\and
Anastasiia Bocharova\inst{1}\orcidlink{0009-0008-1244-3606}
\and
Alexandra Veledina\inst{1,2}\orcidlink{0000-0002-5767-7253}  
}

\institute{
Department of Physics and Astronomy, FI-20014 University of Turku, Finland\\
\email{varpu.a.ahlberg@utu.fi}
\and
Nordita, KTH Royal Institute of Technology and Stockholm University, Hannes Alfv\'ens v\"ag 12, SE-10691 Stockholm, Sweden
}

\date{Received XXXX / Accepted XXXX}

 
  \abstract
   {Cygnus X-3 is the only known Galactic high-mass X-ray binary with a Wolf-Rayet companion.
   Recent X-ray polarimetry results with the Imaging X-ray Polarimetry Explorer have revealed it as a concealed ultraluminous X-ray source.
   It is also the first source where pronounced orbital variability of X-ray polarization has been detected---notably with only one polarization maximum per orbit.}
   {
   Polarization caused by scattering of the source X-rays can only be orbitally variable if the scattering angles change throughout the orbit.
   Since this requires an asymmetrically distributed medium around the compact object, the observed variability traces the intrabinary structures.
   The single-peaked profile further imposes constraints on the possible geometry of the surrounding medium.
   Therefore, the X-ray polarization of Cygnus X-3 is the first opportunity to study the wind structures of high-mass X-ray binaries in detail.
   We aim to uncover the underlying geometry through analytical modeling of the polarized variability.
   Knowledge of these structures could be extended to other sources with similar wind-binary interactions.
   }
   {We study the variability caused by single scattering in the intrabinary bow shock, exploring both the optically thin and optically thick limits.
   We consider two geometries of the reflecting medium, the axisymmetric parabolic bow shock and the parabolic cylinder shock, and determine the geometry that best matches the X-ray polarimetric data.
   }
   {We find that the peculiar properties of the data can only be replicated with a cylindrical bow shock with asymmetry across the shock centerline and significant optical depth.
   This geometry is comparable to shocks formed by the jet-wind or outflow-wind interactions.
   The position angle of the orbital axis is slightly misaligned from the orientation of the radio jet in all our model fits.
   }
   {}


   \maketitle
%

\section{Introduction}

Cygnus X-3 (Cyg X-3) is a peculiar high mass X-ray binary (HMXB) with a Wolf-Rayet (WR) companion \citep{Kerkwijk92}.
The compact object orbits through the dense WR wind, showing infrared and X-ray orbital variability with a period of 4.8 hours \citep{Becklin73, Fender99, Bonnet81}.
The system has a low inclination of about $30\degr$ as measured through its orbital light curve \citep{Antokhin22} and spectrum \citep{Vilhu2009}, as well as its X-ray polarization \citep{Veledina2024NatAs, Veledina2024b}. 
The inclination of its radio outflow was measured in \citet{Miller2004} at a lower value of ${\sim}10\degr$.
The source features distinct X-ray spectral states that are correlated with its radio flux \citep{Szostek2008}.
It is most frequently found in a hard X-ray, quiescent radio spectral state. 
It often transitions to an intermediate X-ray state, sometimes followed by an ultrasoft state and a soft non-thermal state, during which major radio ejections are observed.
Although the spectrum of Cyg X-3 is close to typical hard/soft states, in contrast to other X-ray binaries, it has strong Compton reflection-like features \citep{Hjalmarsdotter2008, Hjalmarsdotter2009, Koljonen2018}.
X-ray polarimetric observations with the Imaging X-ray Polarimetry Explorer (IXPE) provided evidence for the presence of obscuring material close to the compact object \citep{Veledina2024NatAs}.
The average polarization degree (PD) of about 20\% in the hard state and 10\% in the intermediate and ultrasoft states \citep{Veledina2024NatAs, Veledina2024b} indicate the central source is concealed from view.
This suggests that the compact object is surrounded by an optically thick funnel-like outflow, which is anticipated to form an ultraluminous X-ray source \citep{King2001}.

One unusual aspect of the X-ray polarization of Cyg X-3 is its significant orbital variability that remains consistent across spectral states.
The PD varies by about 5\% in all states, while the polarization angle (PA) varies by about \(10 \degr\) in the hard and intermediate states \citep{Veledina2024NatAs} and by \(20 \degr\) in the ultrasoft state \citep{Veledina2024b}.
Orbital variability of X-ray polarization has only been detected in two other sources: low-mass X-ray binary GS 1826$-$238 \citep{Rankin2024} and the widely studied HMXB Cygnus X-1 \citep[Cyg X-1,][]{Kravtsov25}, both of which feature a much smaller amplitude compared to Cyg X-3.
Interestingly, the PD and PA variations of both Cyg X-1 and Cyg X-3 exhibit only one peak per orbit.
In Cyg X-3 in particular, the PD and PA peak at roughly the same orbital phase.
Polarization caused by electron scattering typically produces two peaks per orbit unless significant asymmetry is present \citep{Brown78}.

The orbital light curve of Cyg X-3 has one pronounced minimum, attributed to the superior conjunction \citep{vanderklis81,vankerkwijk93}, and smaller minima indicating presence of one or more sub-structures in the WR wind \citep{Vilhu2009,Antokhin22}.
A secondary flux minimum at the orbital phase of 0.4 was explained by the presence of a ``clumpy trail'': a region of clumped gas caused by a jet bow shock colliding with the wind \citep{Vilhu2013}.
The X-ray and infrared light curves were successfully modeled  with three absorbing components: the wind, the clumpy trail, and the bow shock \citet{Antokhin22}.
All three components might potentially contribute to the polarization light curve as well.
Scattering in the wind will only have a minor effect on the polarization since it features an underlying spherical symmetry and multiple scatterings.
The clumpy trail receives less flux than the bow shock and probably lacks the asymmetry required for producing single-peaked variability.
This leaves the bow shock as the most likely primary cause of the orbital variability.
An optically thick cylindrical bow shock can produce the single-peaked variability at a qualitative level, although quantitative fit was not found \citep{Veledina2024NatAs}.

In this paper, we develop the first model that can quantitatively fit the orbital variability of X-ray polarization in Cyg X-3.
We extract useful parameters from the fit, among them the shape of the bow shock.
In Sect. \ref{sect:model}, we introduce the equations for polarized single scattering and our model geometries;
in Sect. \ref{sect:results}, we perform model fits to compare the different geometries;
in Sect. \ref{sect:discussion}, we discuss the implications of our results; and in Sect. \ref{sect:summary}, we provide a brief summary.


\section{Model setup and data} \label{sect:model}
\subsection{Geometry} \label{sect:polref}

We represent linear polarization with the Stokes flux parameters $F_I$, $F_Q$, and $F_U$.
Using these parameters, the PD and PA can be obtained by $P = \sqrt{F_Q^2 + F_U^2}/F_I^2$ and  $\tan 2\chi = F_U/F_Q$, respectively.
Stokes $F_Q$ and $F_U$ can also be expressed in their normalized form: $q = F_Q/F_I$ and $u = F_U/F_I$.
We assume that the compact object is in a counterclockwise circular orbit characterized by the orbital phase angle, $\varphi$, with $\varphi=0$ corresponding to the superior conjunction.
We use a coordinate system fixed on the compact object, with the orbital axis, $\vec{\hat{\Omega}} = (0,0,1)$, aligned with the $z$-axis and the direction to the companion star aligned with the $y$-axis (see Fig. \ref{fig:geom}).
In these coordinates, the direction towards the observer is
\begin{equation}
    \vec{\hat o} = (\sin i \sin \varphi, \sin i \cos \varphi, \cos i),
\end{equation}
where $i$ is the observer inclination.
The radial vector of a scattering point is
\begin{equation}
    \vec{\hat r} = (\sin \theta \cos \phi, \sin \theta \sin \phi, \cos \theta),
\end{equation}
where $\theta$ is the angle from the $z$-axis and $\phi$ is the azimuth measured from the $x$-axis.
The cosine scattering angle is
\begin{equation}
    \mu = \vec{\hat{r}} \cdot \vec{\hat{o}} = \cos i \cos \theta  + \sin i \sin \theta \sin (\phi + \varphi).
\end{equation}
The normal of the scattering plane is the polarization pseudo-vector:
\begin{equation}
    \vec{\hat p} = \frac{\vec{\hat{o}} \times \vec{\hat{r}}}{|\vec{\hat{o}} \times \vec{\hat{r}}|}.
\end{equation}
The polarization angle depends on the choice of polarization basis, for which we select the projection of the orbital axis on the plane of the sky:
\begin{align}
    \vec{\hat{e}_1} &= \frac{\vec{\hat{\Omega}} - \cos i \vec{\hat{o}}}{\sin i} = (-\cos i \sin \varphi, -\cos i \cos \varphi, \sin i), \\
    \vec{\hat{e}_2} &= \frac{\vec{\hat{o}} \times \vec{\hat{\Omega}}}{\sin i} = (\cos \varphi, -\sin \varphi, 0).
\end{align}
The PA of light scattered from point $\vec{\hat r}$ (assuming the incident light is unpolarized) is
\begin{align}
    \cos \chi &= \vec{\hat{e}_1} \cdot \vec{\hat{p}} = - \frac{\sin \theta \cos (\varphi + \phi)}
    {\sqrt{1 - \mu^2}}, \\
    \sin \chi &= \vec{\hat{e}_2} \cdot \vec{\hat{p}} = \frac{\sin i \cos \theta -
    \cos i \sin \theta \sin (\varphi + \phi)}
    {\sqrt{1 - \mu^2}}.
\end{align}
The polarization basis used in the observations has the $z$-axis aligned with the north-south direction in the sky, so the observed $q$ and $u$ are rotated by the position angle of the orbital axis $\Omega$ (that is measured counterclockwise from the north):
\begin{align}
    q_\mathrm{obs} &= q\cos 2 \Omega - u \sin 2 \Omega,\\
    u_\mathrm{obs} &= q \sin 2 \Omega + u \cos 2 \Omega.
\end{align}

\subsection{Polarized Scattering}

Let \(F_{I,0}(r, \theta, \phi), F_{Q,0}(r, \theta, \phi),\) and \(F_{U,0}(r, \theta, \phi)\) be the Stokes fluxes of radiation incident to an electron scattering envelope as a function of radial distance from the source, $r$, and the angles $\theta$ and $\phi$.
The Stokes vector of light scattered once by an optically thin envelope is \citep{Brown78, Fox93}
\begin{equation} \label{eq:thinintegral}
    \tilde{F}_\mathrm{sc} = 
    \frac{\sigma_\mathrm{T}}{4\pi D^2}
    \int\limits_V n(\vec{r}) \mathbf{P}(i, \pi/2 - \varphi, \theta, \phi)
    \left( \begin{matrix} 
      F_{I,0} (r, \theta, \phi)\\
      F_{Q,0} (r, \theta, \phi)\\
      F_{U,0} (r, \theta, \phi)
    \end{matrix} \right)
    \, \mathrm{d}V,
\end{equation}
where \(\mathbf{P}(\theta_1, \phi_1, \theta_2, \phi_2)\) is the scattering matrix \citep[p. 40-42]{Chandrasekhar60}, \(\sigma_\mathrm{T}\) is the Thomson cross-section, $D$ is the distance to the observer, and \(n(\vec r)\) is the electron number density of the envelope as a function of the radius vector $\vec r$.
The first azimuthal angle of the scattering matrix is set to $\pi/2 - \varphi$ so that its definition matches that of \citet{Chandrasekhar60}.
If the incident emission from the source is unpolarized and isotropic, the integral simplifies to
\begin{equation} \label{eq:simplethinintegral}
    \tilde{F}_\mathrm{sc} = 
    \frac{L}{4 \pi D^2} \frac{3\sigma_\mathrm{T}}{16\pi}
    \int\limits_V \frac{n(\vec{r})}{r^2} 
    \left( \begin{matrix} 
      1 + \mu^2\\
      (1 - \mu^2)\cos 2\chi\\
      (1 - \mu^2)\sin 2\chi
    \end{matrix} \right)
    \, \mathrm{d}V,
\end{equation}
where $L$ is the source luminosity.
The integrand can be separated into components varying at the orbital frequency and its second harmonic:
\begin{align}
    F_{I, \mathrm{sc}} &\propto - \cos(2\varphi + 2\phi) \sin^2 \theta \sin^2 i + \sin(\varphi + \phi) \sin 2 \theta \sin 2 i , \nonumber \\ 
    F_{Q, \mathrm{sc}} &\propto \cos(2\varphi + 2\phi) \sin^2 \theta [1 + \cos^2i] + \sin(\varphi + \phi) \sin 2\theta \sin 2i ,\nonumber \\
    F_{U, \mathrm{sc}} &\propto  \sin (2\varphi + 2\phi) \sin^2 \theta \cos i  - \cos(\varphi + \phi) \sin 2\theta \sin i. \label{eq:QUfourier}
 \end{align}
The orbital frequency components of two scattering points located at $\theta$ and $\pi - \theta$ have the same amplitude with opposite signs.
Thus, scattering from any optically thin envelope that is symmetric relative to the $x$-$y$ plane will only produce the second harmonic of the orbital frequency.
This behavior also applies to polarized or anisotropic incident radiation if its angular distribution shares this symmetry.

For an optically thick surface with uniform scattering albedo $\lambda$, the single-scattered flux is
\begin{equation} \label{eq:thickintegral}
    \tilde{F}_\mathrm{sc} = 
    \frac{\lambda}{4\pi D^2}
    \int\limits_S \mathbf{P}(i, \pi/2 - \varphi, \theta, \phi)
 \left( \begin{matrix} 
      F_{I,0} (r, \theta, \phi)\\
      F_{Q,0}(r, \theta, \phi)\\
      F_{U,0} (r, \theta, \phi)
    \end{matrix} \right)
    \frac{\eta \eta_0}{\eta + \eta_0}
    \, \mathrm{d}S,
\end{equation}
where $\eta = \vec{\hat{o}} \cdot \vec{\hat{n}}$ is the angle between the observer and the surface normal and $\eta_0 = -\vec{\hat{r}} \cdot \vec{\hat{n}}$ is the angle between the normal and the radial vector of the reflecting point.
We treat the optically thick surface as fully opaque; the surface is only visible when $\eta > 0$ and $\eta_0 > 0$. 
We do not consider self-obscuration of the shock.

The total flux is a combination of the shock scattering and direct radiation.
This can be expressed in units of direct flux as
\begin{equation}
    \tilde{F}_\mathrm{tot} = F_0 \left( \begin{matrix} 
      1 + F_{I, \mathrm{sc}}/F_0\\
      q_0 + F_{Q, \mathrm{sc}}/F_0\\
      F_{U, \mathrm{sc}}/F_0
    \end{matrix} 
    \right),
\end{equation}
where $F_0$ and $q_0$ represent a constant component.
We only consider zero $u_0$, which is expected from the constant component that is symmetric relative to the orbital plane.

\subsection{Model for the incident light}\label{sect:incident}

We consider the case of incident flux from the reflecting funnel model described in \citet{Veledina2024NatAs}.
In the model, unpolarized and isotropic incident radiation of the compact object with flux $F_\star = \displaystyle{\frac{L}{4\pi r^2}}$ is reflected off an optically thick cone.
To simplify the problem, we assume that the point the light is reflected towards is distant, so the incident light is a point source for the scattering medium.
The geometry is characterized by two parameters: the grazing angle of the cone, $\alpha$, and the distance from the compact object to the outer edge of the cone, $R$.
The observed average polarization of Cyg X-3 in its hard state is consistent with parameter values of $\alpha = 10 \degr$ and $R=10$ \citep{Veledina2024NatAs}; thus, we use these values in our calculations to reduce the number of free parameters.
We also combine the normalization parameters, including the constant flux $F_0$, the albedo of the cone $\lambda_\mathrm{cone}$, and the albedo of the shock $\lambda_\mathrm{shock}$, into one parameter: $\epsilon = \lambda_\mathrm{cone} \lambda_\mathrm{shock} F_\star/F_0$.

\begin{figure} 
\begin{center}
\includegraphics[width=\linewidth]{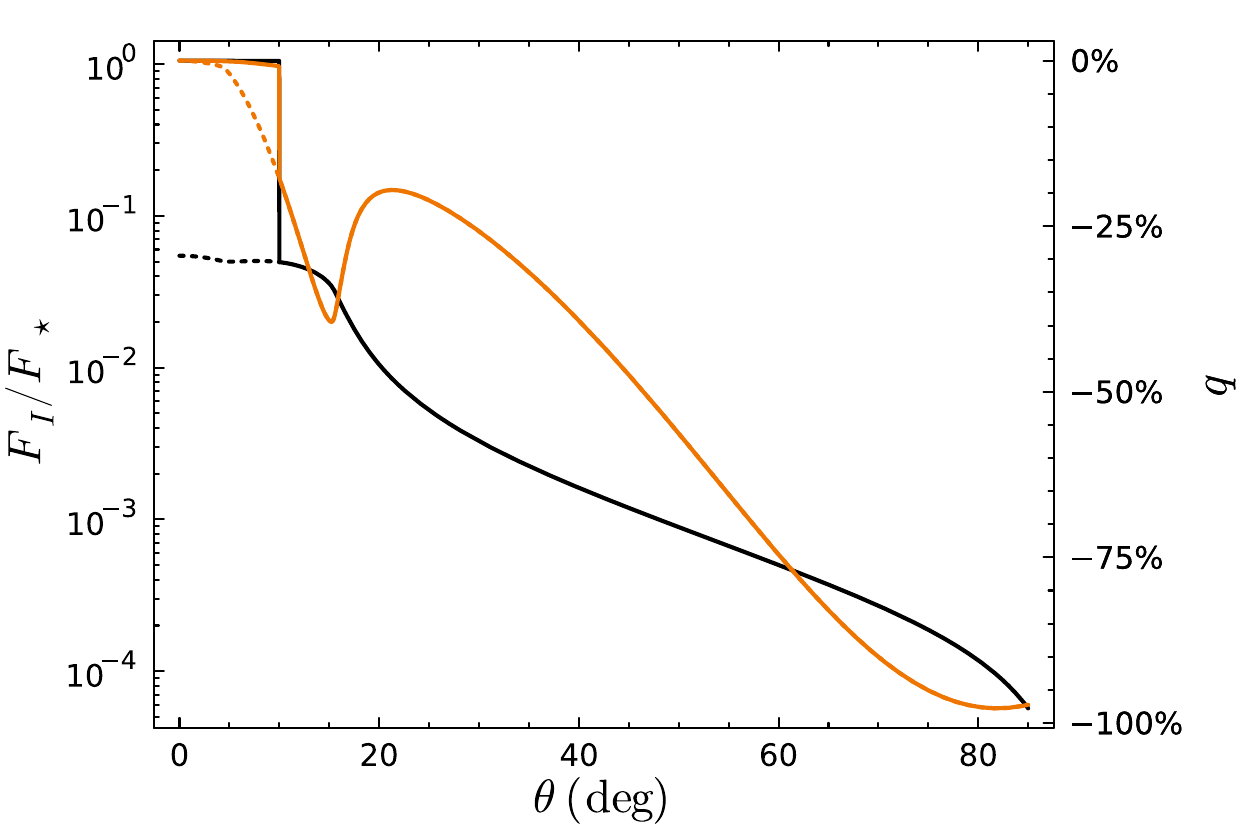}
\end{center}
\caption{Stokes $F_I/F_\star$ (black) and $q$ (orange) of light reflected from an optically thick conical surface as a function of the angle $\theta$. Dotted lines show the reflected light without direct radiation from the source.}
\label{fig:conepol}
\end{figure}

Figure \ref{fig:conepol} shows $F_\mathrm{I,0}$ and $F_\mathrm{Q,0}$ as a function of the inclination, $\theta$, with the single-scattering albedo set to unity.
At angles $\theta < \alpha$, the direct radiation from the central source is visible, and the total flux is about 20 times greater than the reflection alone.
With a more realistic single-scattering albedo of $\lambda \sim 0.1$ for neutral matter, the ratio between the reflected flux and the direct flux at $\theta \sim 45^\circ$ is only about $10^{-4}$.
A highly ionized cone with multiple scatterings provides a similar ratio between the beamed flux and the reflection \citep{Dauser2017}.
Due to the significant obscuration, we only consider cases where the region of the bow shock that reaches above the funnel is sufficiently rarefied that it gives only a minor contribution to the total scattered flux.
Otherwise, scattering from gas directly above the funnel would dominate over all other regions, resulting in nearly constant (orbital phase-independent) polarization.

\begin{figure*} 
\begin{center}
\includegraphics[width=0.27\linewidth]{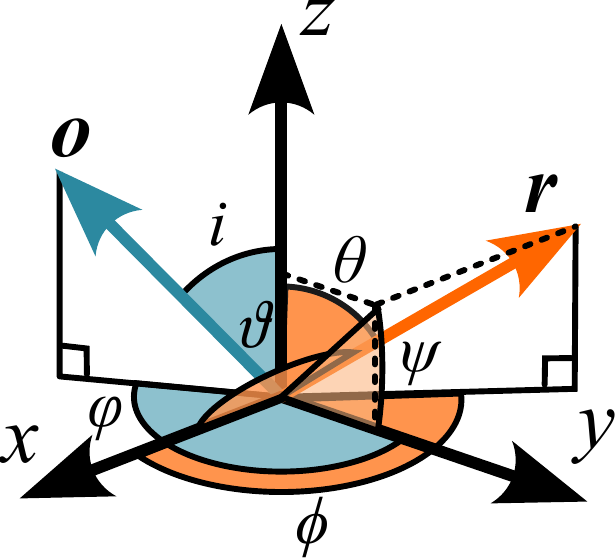}
\includegraphics[width=0.3\linewidth]{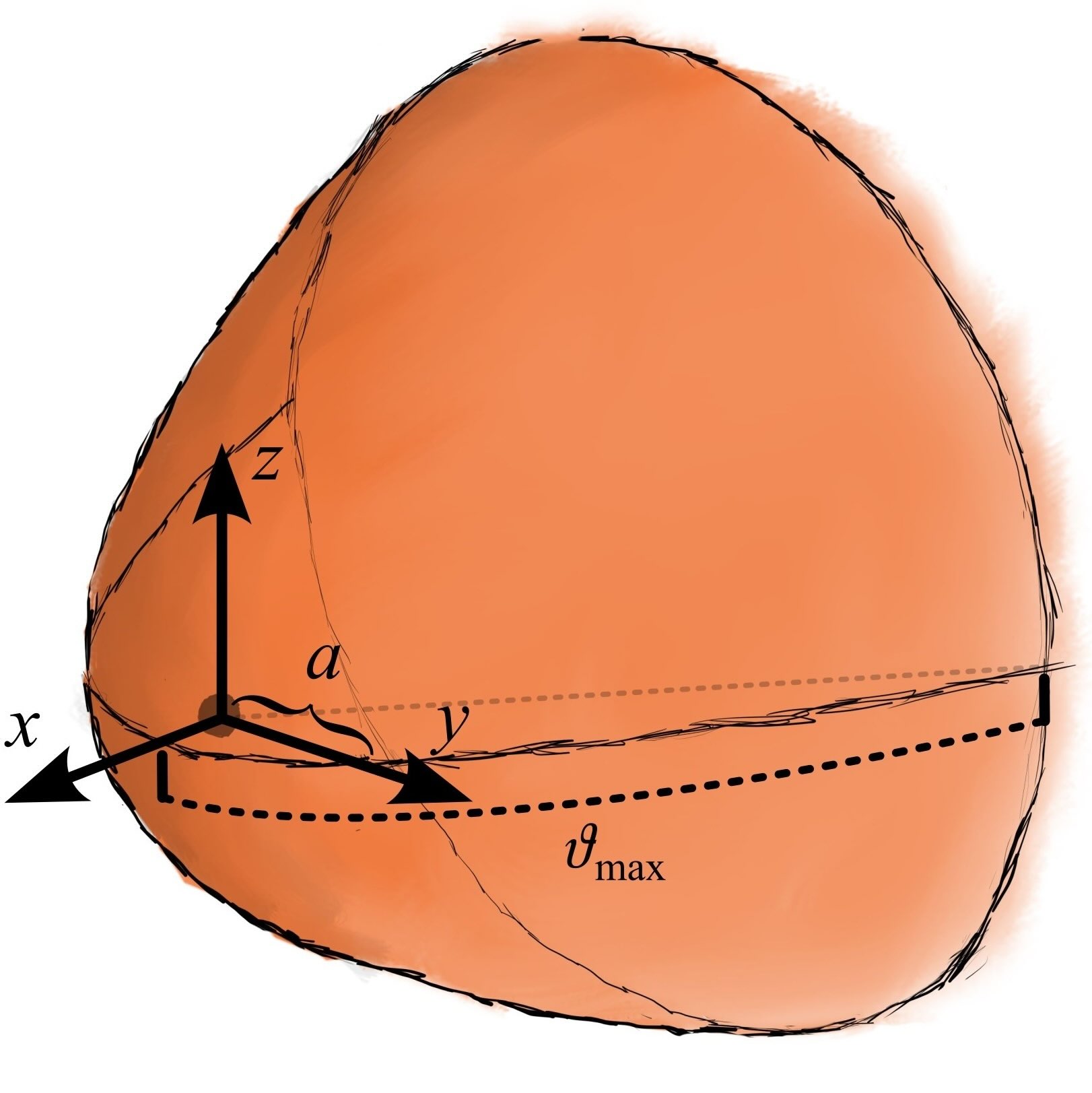}
\includegraphics[width=0.3\linewidth]{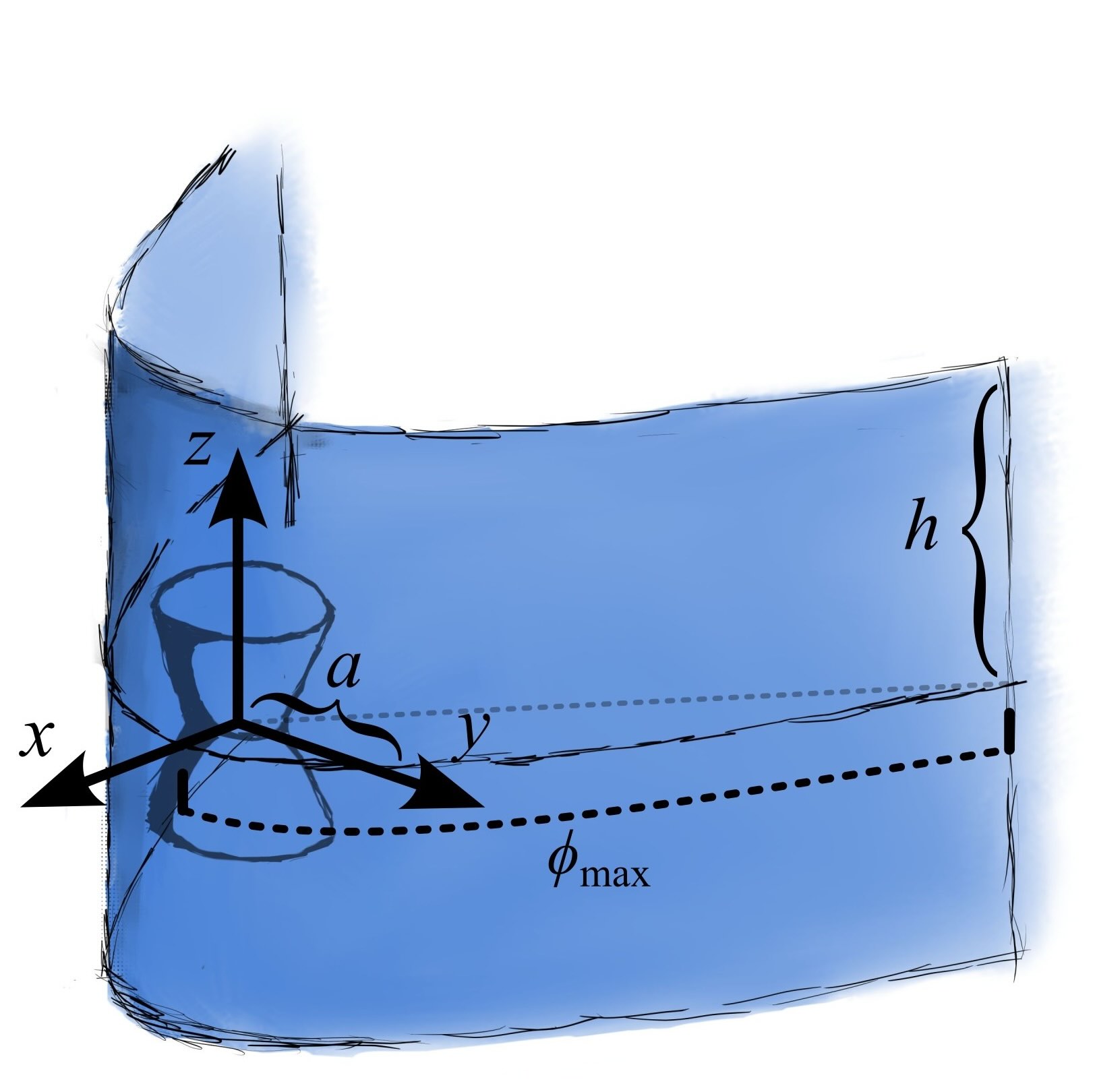}
\end{center}
\caption{Illustration of the coordinate system used in this work (left) and sketches of the parabolic (center) and parabolic cylinder (right) shock geometries.}
\label{fig:geom}
\end{figure*} 

\subsection{Scattering from a point-like cloud}

We first consider the simple case where the shock is represented by a point scattering the incident radiation, located at a fixed inclination, $\theta_0$, and azimuth, $\phi_0$.
For unpolarized incident radiation, the reflected Stokes flux is
\begin{equation} \label{eq:RVM}
    \tilde{F}_\mathrm{sc} = \epsilon F_0 \frac{3}{8} \left( \begin{matrix} 
      1 + \mu^2\\
      (1 - \mu^2) \cos 2\chi\\
      (1 - \mu^2) \sin 2\chi
    \end{matrix} \right),
\end{equation}
where $\epsilon = f_\mathrm{sc} F_\star / F_0$ is the flux normalization factor that includes the scattering fraction, $f_\mathrm{sc}$.
This treatment of scattering approaches Eq.~(\ref{eq:simplethinintegral}) when the size of the scattering region is sufficiently small.
This approach is similar to the phase-dependent polarization in the rotating vector model \citep[used to explain the variable polarization of pulsars,][]{Poutanen2020, Radhakrishnan1969}.

\subsection{Parabolic bow shock} \label{sect:parabolic}

In an idealized case, a spherical object traveling through a medium will produce a parabolic bow shock.
The geometry is illustrated in Fig. \ref{fig:geom}.
Density at the shock boundary for an ideal strong bow shock is \citep{DuPont2024} 
\begin{equation}
    n_\mathrm{sh} = \frac{\gamma +1}{\gamma-1} n_\mathrm{wr},
\end{equation}
where \(n_\mathrm{wr}\) is the density of the Wolf-Rayet wind and $\gamma$ is the adiabatic index.
For the classical result of \(\gamma = 5/3\), the shock is therefore four times as dense as the wind.

We set both the direction of orbital motion and the shock apex along the $x$-axis.
To model a misaligned shock, the orbital phase angle can be shifted by the angle between the directions of the shock apex and the orbital motion.
The shape of the parabolic bow shock is defined by the relation
\begin{equation}
    x = 1 - y^2/a^2 - z^2/a^2,
\end{equation}
where $a$ is the radius of the shock cross-section at $x=0$ in units of distance to the shock apex.
Due to the axial symmetry of the shock around $x$, it is easier to perform the calculations in rotated spherical coordinates measuring the inclination angle with respect to the $x$-axis, $\cos \vartheta = \sin \theta \cos \phi$, and the azimuthal angle from the $y$-axis toward the $z$-axis, $\tan \psi = \cos \theta / (\sin \theta \sin \phi)$.
We parameterize the extent of the shock by the maximal angle from the apex, $\vartheta_\mathrm{max}$.
The distance from the central source to the parabolic surface is
\begin{equation}
    r_\mathrm{p} = \frac{\sqrt{\cos^2 \vartheta + 4 a^{-2} \sin^2 \vartheta  } - \cos \vartheta}{2 a^{-2} \sin^2 \vartheta},
\end{equation}
and the normal to the surface of the parabola is
\begin{equation}
    \vec{\hat n} = \frac{(-2r_\mathrm{p} \sin \vartheta \cos \psi, -2r_\mathrm{p}  \sin \vartheta \sin \psi, -a^2)}{\sqrt{a^4 + 4r_\mathrm{p}^2  \sin^2 \vartheta}}.
\end{equation}

To model an optically thin shock using the volume integral in Eq. (\ref{eq:thinintegral}), we treat the shock as a geometrically thin shell with constant density:
\begin{equation}
    n(r) = n_\mathrm{sh} \delta(r - r_\mathrm{p}).
\end{equation}
Using this notation, we can write the surface element of the parabola for the integral in Eq. (\ref{eq:thickintegral}) as
\begin{equation}
    \mathrm{d}S = r_\mathrm{p}^2 \sqrt{\sin^2 \vartheta + \left( \frac{\sin \vartheta}{r_\mathrm{p}} \frac{\partial r_\mathrm{p}}{\partial \vartheta}\right)^2 + \left( \frac{1}{r_\mathrm{p}} \frac{\partial r_\mathrm{p}}{\partial \psi} \right)^2} \mathrm{d} \vartheta \, \mathrm{d} \psi.
\end{equation}

\subsection{Cylindrical bow shock} \label{sect:cylindrical}

To account for the potential modification of the shock shape caused by the jet-wind or funnel-wind interaction, we consider the shock as a sector of a parabolic cylinder as depicted in Fig. \ref{fig:geom} \citep[blue surface,][]{Yoon15, Yoon16}.
With the apex along the $x$-axis, the cylinder is defined by the equation
\begin{equation}
    x = 1 - y^2/a^2,
\end{equation}
where $a$ is the radius of the cylinder at $x=0$.
A misaligned shock can again be modeled by shifting the orbital phase angle.
The maximal extent of the shock is defined by the azimuthal angle $\phi_\mathrm{max}$.
The cylindrical radius $\rho = \sqrt{x^2 + y^2}$ of the shock surface is
\begin{equation}
    \rho_\mathrm{pc} = \frac{\sqrt{\cos^2 \phi + 4a^{-2} \sin^2 \phi} - \cos \phi}{2 a^{-2} \sin^2 \phi},
\end{equation}
and the surface normal is
\begin{equation}
    \vec{\hat n} = \frac{(-a^2, -2 \rho_\mathrm{pc} \sin \phi, 0)}{\sqrt{a^4 + 4 \rho_\mathrm{pc}^2  \sin^2 \phi}}.
\end{equation}
The integral in Eq. (\ref{eq:thickintegral}) is most conveniently performed in cylindrical coordinates, where the surface element is
\begin{equation}
    \mathrm{d}S = \sqrt{\rho_\mathrm{pc}^2 + \left(\frac{\partial \rho_\mathrm{pc}}{\partial \phi} \right)^2} \mathrm{d} \phi \, \mathrm{d} z.
\end{equation}


\subsection{Data analysis}

We use phase-resolved polarimetric data from IXPE \citep[observation ID 02001899][]{Weisskopf2022}. 
It is a NASA/ASI mission dedicated to polarimetric observations in the energy range 2--8 keV. It includes three X-ray telescopes each comprised of a Mirror Module Assembly and a polarization sensitive gas-pixel detector unit \citep{Baldini2021, Soffitta2021}. 

The data were downloaded from the HEASARC public archive\footnote{\url{https://heasarc.gsfc.nasa.gov/docs/archive.html}} and processed with the {\sc ixpeobssim} package version 31.0.1 \citep{Baldini2022} using the detector responses from CalDB version 13. The source photons were extracted from a circular region centered on the source, with a radius of 60\arcsec. The observation was folded with the quadratic ephemeris of \citep{Antokhin2019} using the \texttt{xpphase} method. The data were then grouped in 10 phase bins, and the \texttt{pcube} method was applied to each of the phase bins in the energy range of 3.5--6 keV.
This energy range has little contribution from multiple scatterings or fluorescence, which would reduce the PD as compared to single scattering \citep{Veledina2024NatAs}.

We employed the statistical modeling package Turing.jl, implementing the No-U-Turn sampler \citep{turingjl, NUTS} to perform Bayesian Markov chain Monte Carlo fitting of the data.
This approach aims to find the posterior probability distributions of the model parameters with random sampling.
We calculated the posterior likelihood of each sampled parameter set by comparing the model to the observed Stokes $q$ and $u$, with the assumption that the data are distributed normally with a standard deviation equal to their error.
To speed up calculation of our models that involve numeric integration, we precalculated them as parameter lookup tables and performed linear interpolation to find the results.
We note that the statistically significant inference of all geometric parameters is unrealistic due to the low number of data points.
Consequently, we fixed the inclination to $i=30 \degr$ in all models, and fixed the parameters $a$ and $h$ whenever necessary.
This introduces a systematic uncertainty to the inferred parameters, although the fitting procedure can demonstrate whether a particular shock geometry can replicate the observations.
The specific inclination was chosen since it is consistent with previously measured values \citep{Antokhin22, Vilhu2009, Miller2004}, and we found that varying it by $\pm 10 \degr$ did not significantly alter of parameters of the fit. 


\section{Results} \label{sect:results}

\subsection{Point-like scattering}

We first applied the point-like single scattering model to the data.
Scattering from a point-like cloud can only produce single-peaked orbital variability if it is elevated high above the orbital plane.
This is shown by Eq. (\ref{eq:QUfourier}): the orbital frequency component is $\propto \sin 2\theta_0$, whereas the second harmonic is $\propto \sin^2 \theta_0$, and therefore the single-peaked variability can only dominate at small $\theta_0$.
We assume that the scattering point is above the funnel and receives the direct unpolarized emission from the central source.
The phase of the superior conjunction is not necessarily the observed zero phase, so we sum the phase shift and $\phi_0$ to create a new parameter, $\varphi_0$.
Thus, the model has five parameters in total: $\theta_0$, $\varphi_0$, $\epsilon$, $\Omega$, and $q_0$.

\begin{figure}
\begin{center}
\includegraphics[width=\linewidth]{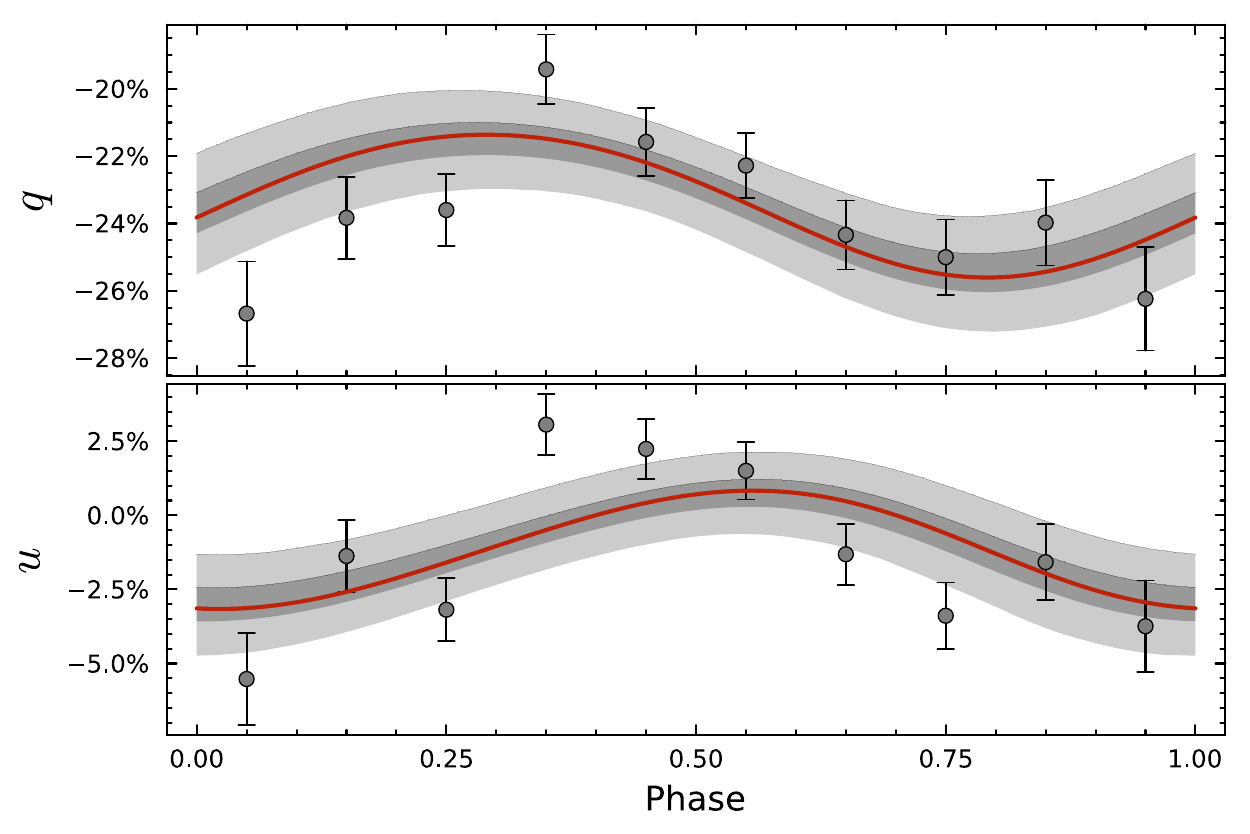}
\end{center}
\caption{Observed Stokes $q$ and $u$ of Cyg X-3 compared to the best fit of the point-like cloud model (dark red) and the 1- and 3-$\sigma$ quantiles of the fit posterior (dark and light regions, respectively).}
\label{fig:dotfit}
\end{figure}

Figure \ref{fig:dotfit} shows the best fit of the model, and the posterior distributions are shown in Fig. \ref{fig:dotcorner}.
The point-like cloud only roughly fits the data with its mostly sinusoidal variability.
This simple model cannot replicate the data in which the peaks of Stokes $q$ and $u$ appear near the same phase.
The parameters $q_0$ and $\epsilon$ are strongly degenerate in this case, and neither can be constrained.
The inclination of the cloud, $\theta_0 = 3\fdg6^{+1\fdg8}_{-0\fdg8}$, places it almost directly above the funnel. 
With the combined phase shift and azimuthal angle of $\varphi_0 = -106\degr\pm11\degr$, the cloud is pointed away from the companion star.
The orbital axis is very closely aligned with the north-south direction, with a position angle of $\Omega = 1\fdg4^{+0\fdg4}_{-0\fdg5}$.
We conclude that the point-like single scattering cloud cannot explain the observed variability, as the PD and PA maxima cannot appear near the same phase.

\subsection{Parabolic bow shock}

\begin{figure}
\begin{center}
\includegraphics[width=\linewidth]{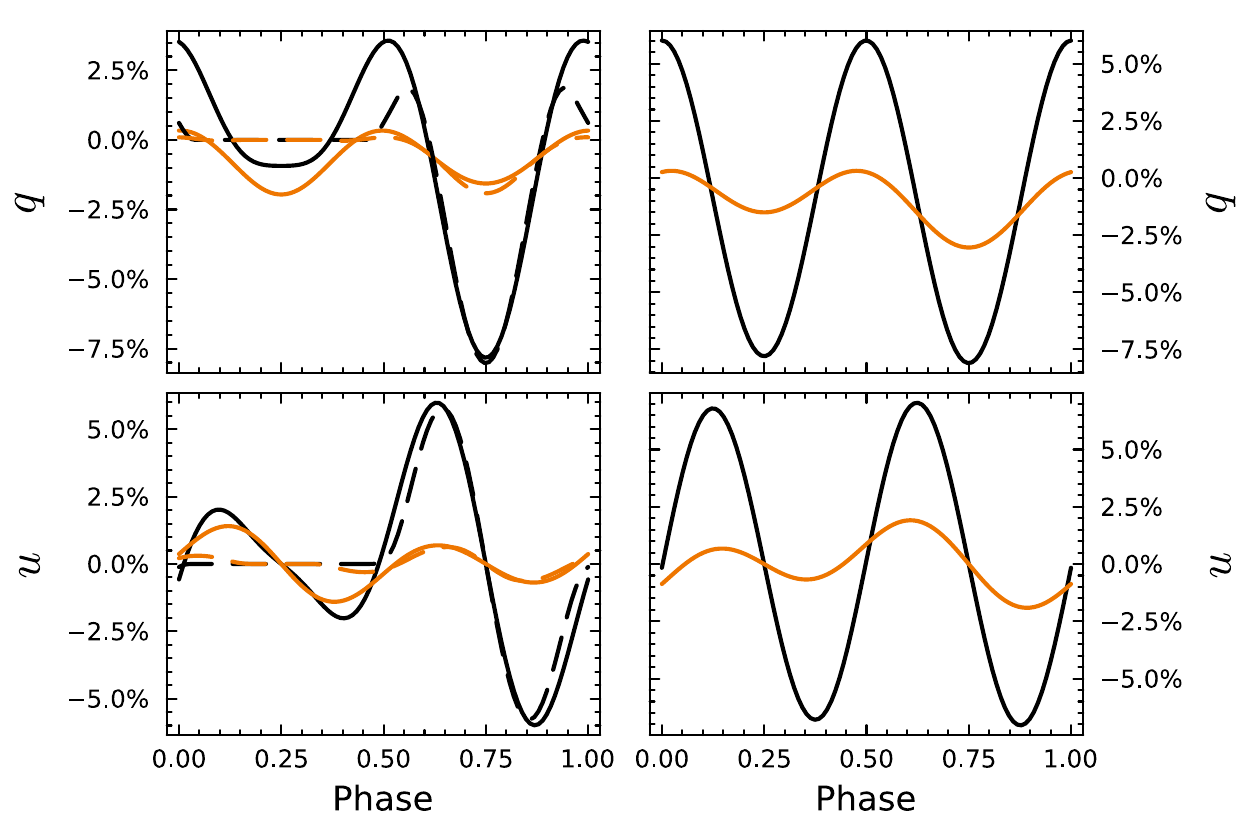}
\end{center}
\caption{Stokes $q$ and $u$ of a theoretical parabolic bow shock of width $\vartheta_\mathrm{max} = 30\degr$ (black) and $\vartheta_\mathrm{max} = 80\degr$ (orange) with $a = 1$ (solid) and $a=3$ (dashed), for both the optically thick (left) and optically thin (right) cases. The flux was normalized so that the maximum of the scattered flux is one tenth of the direct flux.}
\label{fig:parabolic}
\end{figure}

The parabolic bow shock model suffers from the same shortcomings as the point-like cloud.
Constant polarization dominates if the shock covers the funnel opening, and in the optically thin scenario, the axial symmetry would lead to the second harmonic dominating (two peaks per orbital period).
As for the optically thick scenario, we find that the shock would have to be so narrow that it would resemble a wall, which behaves qualitatively the same as a narrow cylindrical shock.
Due to these severe problems, we did not perform any formal fitting of the model to the data.
As an illustration, we calculated the polarized scattering from a narrow parabolic bow shock with different parameters.
In the optically thin case, we introduced asymmetry across the orbital plane by excluding the lower half of the shock, which can be interpreted as an accretion disk obscuring it from view.
For the incident radiation, we used the funnel reflection model described in Sect. \ref{sect:incident}.
The results of our calculations are shown in Fig. \ref{fig:parabolic}.
Even with the added vertical asymmetry, the variability in the optically thin case is dominated by the second harmonic, making fully optically thin shocks unable to match the data.
This is a consequence of the narrow shock lying close to the orbital plane.
The obscuration of the scattering surface is thus necessary for the shock to produce only one polarization peak per orbit.

In the optically thick case, scattered flux from the inner shock surface can be entirely obscured over a range of orbital phases if $a$ is sufficiently large.
In contrast, the optically thin model is entirely unaffected by the parameter $a$.
Wider shocks have lower polarized flux since the range of scattering angles is larger and thus would require a larger scattering fraction to match the observed PD.

\subsection{Symmetric cylindrical bow shock}

\begin{figure}
\centering
\includegraphics[width=\linewidth]{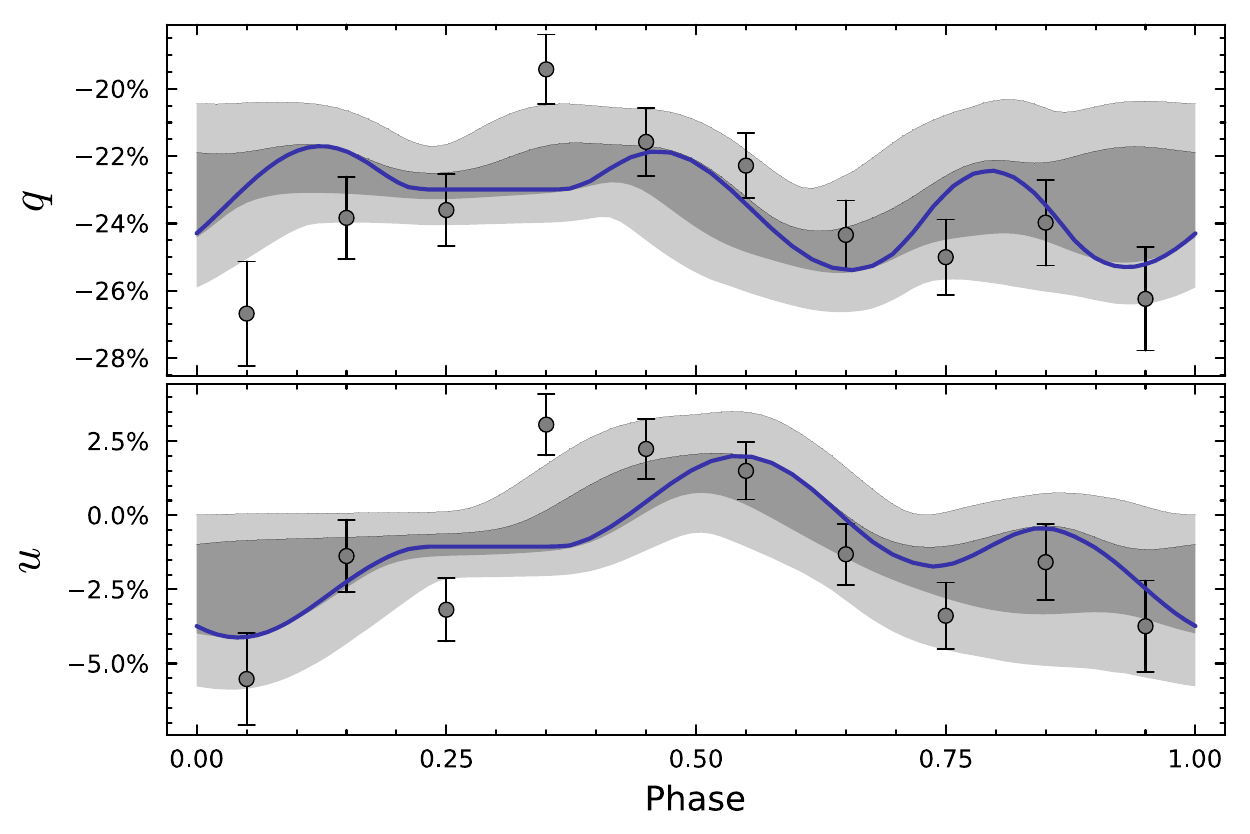}
\caption{Same as Fig. \ref{fig:dotfit} but for the parabolic cylinder shock model.}
\label{fig:symfit}
\end{figure}

Next, we consider the cylindrical bow shock (Fig.~\ref{fig:geom}, right).
We assume an optically thick shock with incident light coming from the cone reflection.
An optically thin cylindrical shock is similar to the parabolic bow shock with its mostly double-peaked variability.
Since this model has too many parameters to infer from the data, we fixed the geometric parameters $a=1.0$, $h=1.0$, and $i=30\degr$.
We tested other values for $a$ and $h$ between about $0.5-2$ and $0.5-5$, respectively, but could not find a significantly better match.
The angle between the shock apex and the direction of motion is not separable from the phase shift of the observation, so we again combine them into a single parameter, $\varphi_0$.
This leaves us with five parameters to fit: $\phi_\mathrm{max}$, $\epsilon$, $\Omega$, $\varphi_0$, and $q_0$.
Figure \ref{fig:symfit} shows the best fit of the model, and the posterior distributions of its parameters are given in Fig. \ref{fig:symcorner}.
The model follows the observed Stokes $u$ reasonably well but fits $q$ poorly.
This is a consequence of the symmetry of the shock relative to the $x$-$z$ plane.
When the apex of the shock is facing the observer and obscuring the reflecting inner surface, the observer only sees the constant component.
The maxima of Stokes $q$ occur before and after this plateau, and the symmetry of the shock leads to those peaks being identical.
The observed Stokes $q$ only has one clear maximum, which cannot be reproduced by this model.

The posterior distribution of $\varphi_0$ has two preferred values around ${\sim}40 \degr$ and ${\sim}-30 \degr$, with the former being closer to the expected alignment angle of the shock of about $35\degr$ \citep{Antokhin22}.
The posterior distribution of the shock width is similarly double-peaked but centered around a narrow angle of ${\sim}75 \degr$.
The constant component is highly constrained at $q_0 = -22.9\% \pm 0.4\%$, which is well within the error bars of the phase-average polarization in the 3.5--6 keV band \citep{Veledina2024NatAs}.
The flux normalization factor of $\epsilon = 2093^{+480}_{-451}$ is large, corresponding to a peak reflected flux of about $0.1 F_0$.
The analytical model for cone scattering at $i=30 \degr$ predicts a value of $\lambda_\mathrm{cone} F_\star/F_0 \approx 3 \times 10^{2}$, but this value for $\epsilon$ is larger by an order of magnitude.
This can be explained by wind absorption towards the observer or the optically thick single-scattering model underestimating the reflected flux.
The alignment angle of the orbital axis, $\Omega = 1\fdg2 \pm 0\fdg5$, is consistent with the point-like cloud.

\subsection{Asymmetric cylindrical bow shock}

\begin{figure} 
\centering
\includegraphics[width=\linewidth]{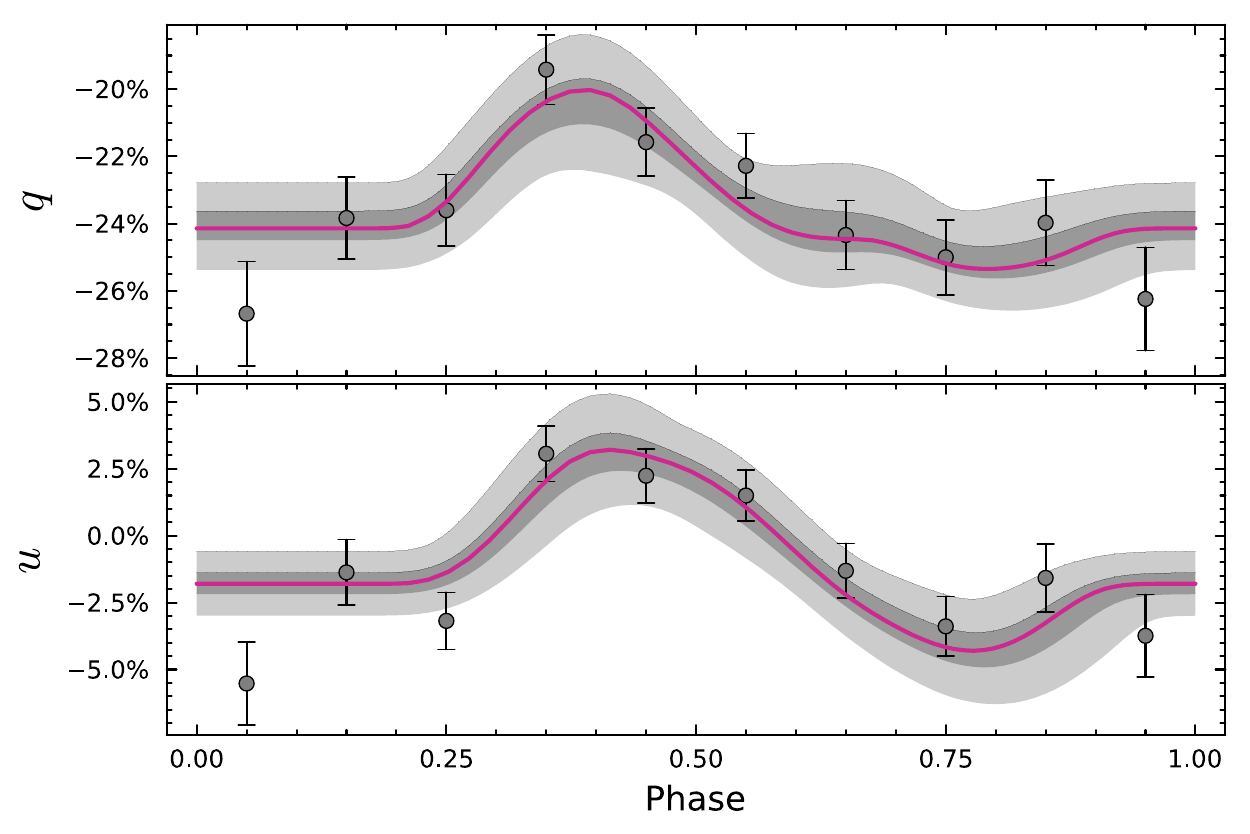}
\caption{Same as Fig. \ref{fig:dotfit} but for the asymmetric parabolic cylinder shock model.}
\label{fig:asymfit}
\end{figure}

We further tested the parabolic cylinder geometry by adding asymmetry across the shock centerline.
In a more accurate model, the shock curves into a spiral shape as it collides with the wind.
However, this is a difficult geometry to model analytically.
Thus, to explore the impact of this asymmetry, we separated widths of the shock on either side of the shock apex into free parameters.
Firstly, $\phi_\mathrm{max1}$ is the width of the shock away from the companion (negative $y$-direction), and secondly, $\phi_\mathrm{max2}$ is the width towards the companion (positive $y$-direction).
As with the symmetric shock, we fixed the parameters $a=1/0.6$, $h=3.0$, and $i=30 \degr$.
The fit was more sensitive to the choice of the shock height and width than the symmetric model, so we tried different values until we found something that could match the amplitude.
The results can be seen in Figs. \ref{fig:asymfit} and \ref{fig:asymcorner}.

\begin{figure}
\begin{center}
\includegraphics[width=\linewidth]{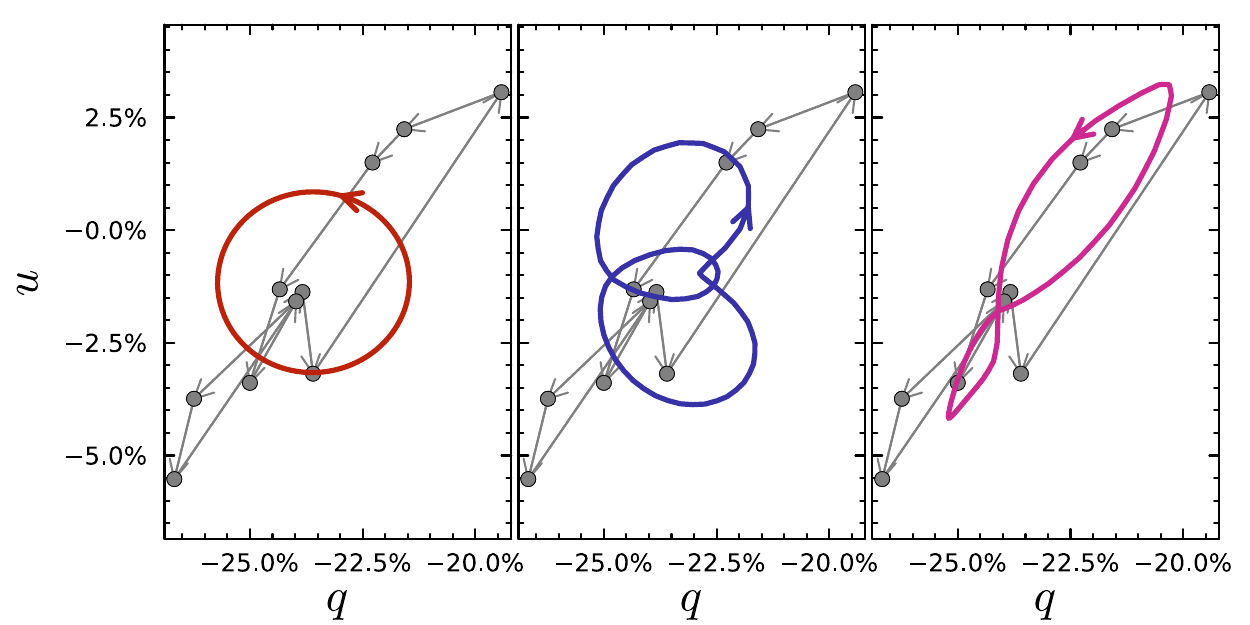}
\end{center}
\caption{The observed variability in the $q$-$u$ plane compared to the best fits of the point-like cloud (left), symmetric cylinder shock (center), and the asymmetric cylinder shock (right). Error bars have been omitted for clarity.}
\label{fig:quplane}
\end{figure}

We find that the model replicates the single peaks of $q$ and $u$ that occur roughly at the same phase.
This is easier to visualize in the $q$-$u$ plane shown in Fig. \ref{fig:quplane}, where the data trace a narrow, skewed shape.
Out of the three model fits, the asymmetric cylinder shock is the only one capable of reproducing the data topology.
However, this model exhibits constant polarization over a relatively large portion of the orbit and thus doesn't fit the polarization near phase zero well.
To quantitatively compare the model fits, we calculated their Akaike information criterion (AIC) values and performed Anderson-Darling (AD) tests comparing the distribution of their normalized residuals to the standard normal distribution.
For the best-fitting point-like, symmetric, and asymmetric shock models, we found $AIC=104$, $AIC=109$, and $AIC=78$, with AD $p$-values of 0.018, 0.011, and 0.56, respectively, favoring the asymmetric model.

Interestingly, the two widths of the shock are starkly different, with the fit preferring a larger $\phi_\mathrm{max1} = 156\degr \, ^{+16 \degr}_{-19 \degr} $ for one side and a much smaller $\phi_\mathrm{max2} = 18\degr \,^{+9\degr}_{-8\degr} $ for the other.
The angle $\varphi_0 = 90\degr \, ^{+7\degr}_{-8\degr}$ places the apex of the shock facing the companion star.
With these parameters, the shock is practically a curved wall facing towards the direction of orbital motion with a small "hook" on the other side of the apex (see magenta shape in Fig. \ref{fig:shockabove}).
This makes the reflection only visible around phase 0.5.

The constant polarization is again strongly constrained at $q_0 = -24.1 \% \pm 0.4 \%$, which is slightly higher than the average polarization.
The flux normalization of $\epsilon = 1967^{+337}_{-317}$ is comparable to the symmetric parabolic cylinder shock, although the reflected flux is larger due to the increased shock height, peaking at $0.4 F_0$.
The position angle $\Omega = 2\fdg1 \degr \pm 0\fdg5 \degr$ is again consistent with the constraints from the other two models.

\section{Discussion} \label{sect:discussion}

\subsection{Properties of the shock}

\begin{figure} 
\centering
\includegraphics[width=0.75\linewidth]{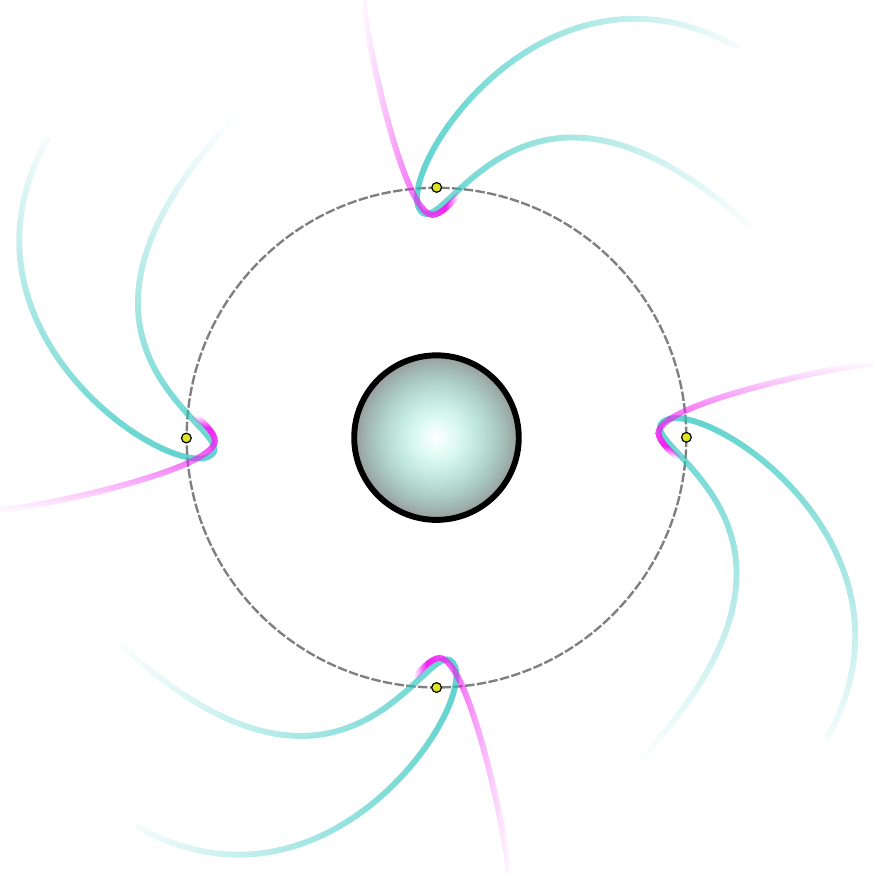}
\caption{Illustration of how the best fit of the asymmetric cylinder shock (magenta) compares to a shock curved by the wind (cyan).}
\label{fig:shockabove}
\end{figure}

The geometry of the intrabinary bow shock in Cyg X-3 is strongly constrained by the X-ray polarimetric data.
If the shock is the primary cause of the orbitally variable polarization, it must be both cylindrical and asymmetric with regard to the shock apex.
This is consistent with theoretical predictions of the bow shock shape in simulations, as the tail of the shock is expected to curve under orbital motion \citep{Eichler93}.
Figure \ref{fig:shockabove} shows how this shape compares to the best fit of our model.
The simplified asymmetric shock is roughly analogous to the curving shock, with the narrow $\phi_\mathrm{max2} = 18\degr \,^{+9\degr}_{-8\degr} $ corresponding to the convex part of the shock.
The surface of this convex side would not be exposed to the central source far from the shock apex, so the extent of the shock surface would appear narrower on that side.
The much wider $\phi_\mathrm{max1} = 156 \degr \, ^{+16 \degr}_{-19 \degr}$, however, does not match the concave half of the curving shock far away from the apex.
A more accurate model would include the physical thickness of the shock, its density distribution, its curvature further away from the shock apex, and multiple scatterings, but this is beyond the scope of this work.

The polarized variability having one peak per orbit can only be recreated if the shock scattering is obscured during certain orbital periods.
Therefore, the shock must be at least moderately optically thick.
Our simplified model assumes an infinite optical thickness and thus cannot constrain the optical depth of the shock.
A model fit of an axisymmetric bow shock in \citet{Antokhin22} determined the shock optical depth through the apex as $\sim1.1$, with the maximum optical depth through the line of sight being $\sim 0.6$.
Our finding that the shock is somewhat optically thick is in line with this model.

We examined the orbital variability only in the hard state, yet it is present in all observations.
The ultrasoft and intermediate states have a smaller constant PD that can be explained with scattering from an optically thin medium above the funnel \citep{Veledina2024b}.
In this scenario, the primary source of X-rays incident on the shock would be the scattered flux rather than the cone reflection. 
Furthermore, the radio jet is quenched in the ultrasoft state, which can impact the shock geometry.
These differences likely explain the changes in the orbital polarization.
Even so, the persistent single-peaked profile makes a semi-stable source for the variability a necessity.
The vertically extended shock therefore has to be sustained by the outflow-wind interaction if the jet is absent.

\subsection{Alignment of the orbital axis and the jet}

The only parameter consistent in all our model fits was the small positive value of the position angle, $\Omega$.
Although the small rotation has little impact on $q$, it shifts $u$ towards negative values.
If the average value of $u$ is slightly negative, it is reasonable to expect fits of different models to find similar values for the position angle.
The near north-south alignment of the orbital axis is consistent with radio observations of the jet \citep{Marti00, Miller2004, Yang23}.
\citet{Yang23} measured the position angle of the innermost radio jet as $-3\fdg1 \pm 0\fdg4$, which differs by a few degrees from our inferred value of $2\fdg1 \pm 0\fdg5$.
While the result is model-dependent, it hints at a small misalignment of the jet with the orbital axis at the scale of the radio observations.

We did not account for a possibly inclined jet and outflow.
However, the orbital gamma ray variability and certain radio observations can be explained with an inclined and precessing jet \citep{Dubus2010}.
The variability can also be explained by a jet bent by the wind, so the inclined jet cannot be confirmed nor rejected \citep{Dmytriiev24}.
The misalignment found by our analysis is significantly smaller than the ${\sim}40\degr$ jet inclination or bending angle indicated by the gamma ray model fits.
It is more in line with the predicted small bending angle that, nonetheless, depends on poorly understood jet parameters \citep{Dmytriiev24}.
The effects of an inclined jet on the orbital variability would be very complicated, as it introduces an orbital phase dependency on the jet-wind interaction \citep{Yoon15}.
We therefore leave this analysis for future work.

\subsection{Impact of other wind structures}

Scattering in the Wolf-Rayet wind contributes to the total observed polarization.
With a spherically symmetric wind, the polarized flux would be near constant, yet the wind is warped by the gravity and the jet-wind interaction.
However, the optically thin wind would primarily produce two PD and PA peaks per orbit and thus cannot be a dominant component in the observed variability \citep{Veledina2024NatAs}.
It might still contribute to the constant Stokes $q_0$, which would be difficult to separate from the constant component caused by the funnel reflection.
Inferring the geometrical parameters of the funnel could be more complicated if the constant polarization is indeed a sum of multiple components.
It is difficult to estimate the contribution from the wind, but simulations of single-scattering in the HMXB wind at lower wind mass loss rates than Cyg X-3 show negligible polarized flux outside of eclipses \citep{Kallman15}.
The higher wind density of Cyg X-3 would lead to more pronounced wind scattering component contribution to total flux, but the presence of multiple scatterings would decrease the polarization degree.

Previous works analyzing the X-ray light curves of Cyg X-3 have assumed a constant initial flux that is modulated at the orbital period by absorption at multiple wind structures \citep[e.g.,][]{Vilhu2013,Antokhin22}.
This analysis neglected the contribution of a variable flux caused by scattering.
Since strong orbital variability of polarization requires that a sizable fraction of the flux is scattered, the light curve should be revisited with this new context.

\begin{figure} 
\centering
\includegraphics[width=\linewidth]{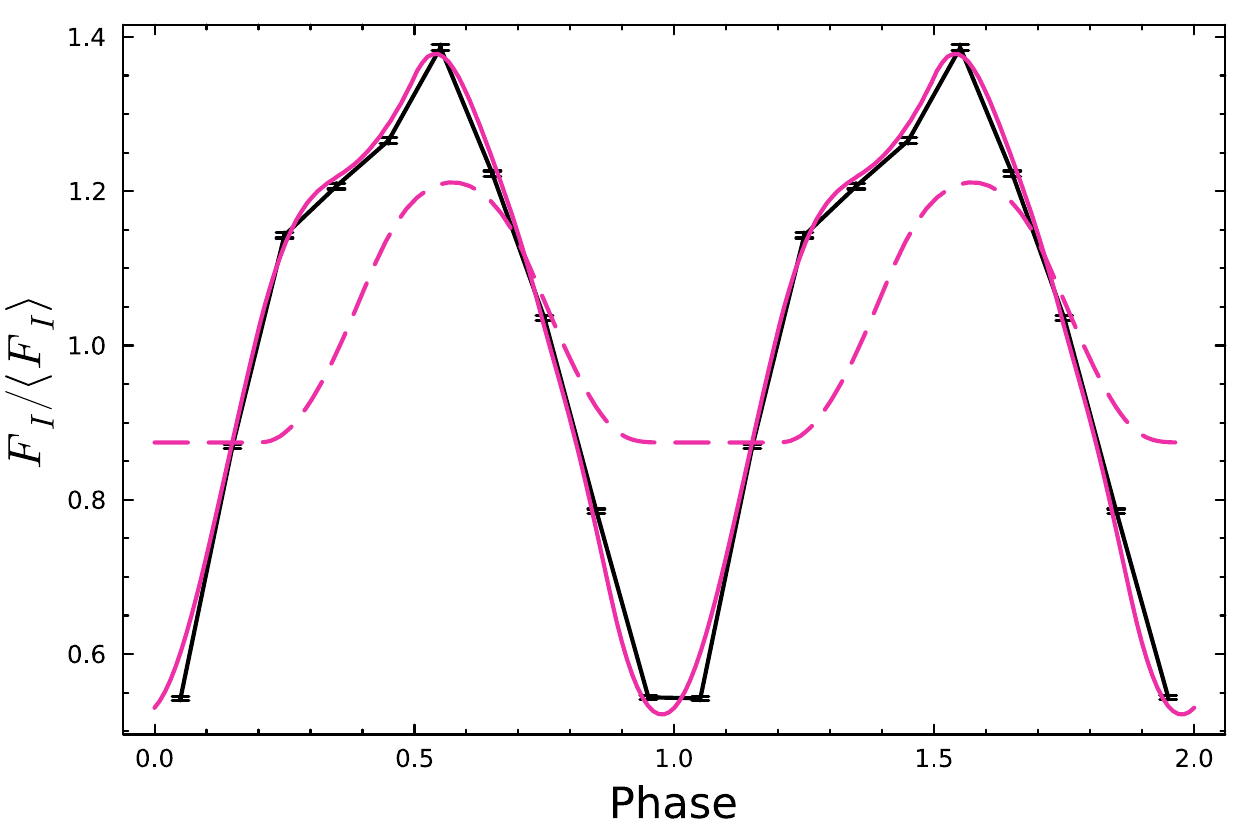}
\caption{Normalized IXPE orbital light curve of Cyg X-3 in the 3.5-6 keV range (black) with the raw reflected flux from the best-fitting asymmetric cylinder shock model (magenta, dashed) and a hand fitted model light curve with wind and clump absorption (magenta, solid).}
\label{fig:Ifit}
\end{figure}

The amount of reflected flux in the parabolic cylinder model is strongly dependent on both $h$ and $a$, and neither parameter is statistically constrained.
Nevertheless, we matched the IXPE light curve by combining the best fit of the asymmetric cylindrical bow shock and the absorption models of \citet{Antokhin22}.
We fixed the inclination to $i = 30 \degr$ and the radius of the WR star to one third of the orbital separation.
The results of our manual fitting are shown in Fig. \ref{fig:Ifit}.
Using the parameter definitions of the aforementioned paper, we used wind absorption with $\tau_0 = 0.33$ and $\phi_0 = 0.24$, and two clump absorbers with $\tau_\mathrm{ct, 1} = 0.35$, $\tau_\mathrm{ct, 2} = 0.25$, $\phi_\mathrm{ct, 1} = 0.47$, $\phi_\mathrm{ct, 2} = 0.69$, $\Delta \phi_\mathrm{ct, 1} = 0.27$, and $\Delta \phi_\mathrm{ct, 2} = 0.19$.
The two absorbers occur before and after the peak of the reflected flux, and in the context of this manual fit, they can be interpreted as absorption in the curving tails of the bow shock (see Fig. \ref{fig:shockabove}).
Absorption in the bow shock itself is not needed in the fit, perhaps because our model already has reduced normalized flux when the shock apex is facing the observer.
Hence, the dimming due to absorption is degenerate with the enhancement of flux due to scattering from the inner surface of the shock.
While this model gives the essence of the observed flux variability, detailed study of the parameter space can be meaningful for the combined set of X-ray and IR light curves, similar to the analysis presented in \citet{Antokhin22}.

Our assumption that the gas above the outflow must have negligible optical depth is complicated by the jet-wind interaction.
Specifically, the wind causes the jet to curve, and so the bow shock would eventually reach above the funnel.
The jet in Cyg X-3 is not strongly bent until a recollimation shock forms about one orbital separation above the orbital plane \citep{Yoon16}.
This is orders of magnitude more distant from the source than the shock apex, so the scattered flux from the lower shock would at least be comparable to the scattered flux from high above.
Assuming that the wind density decreases away from the WR star with the inverse square of distance, the wind near the recollimation shock would thus be about half as dense as the wind near the compact object. 
The part of the bow shock exposed to the direct radiation is therefore significantly less dense than the lower parts of the shock.
All things considered, it is safe to conclude that scattering from the bow shock far above the outflow does not dominate.

\subsection{Comparison with Cyg X-1}

Similar bow shock structures should be present in other HMXBs, yet Cyg X-3 remains the only one with pronounced orbital X-ray polarization.
In particular, Cyg X-1 is another wind-fed microquasar that was observed with IXPE, but the initial analysis by \citet{Krawczynski22} found no statistically significant orbital variability.
Subsequent analysis of the large set of observations showed that orbital variability with a single polarization maximum per orbit is present in Cyg X-1, albeit with PD amplitude of $\sim2\%$ \citep{Kravtsov25}. 
The lower statistics of the orbital phase resolved polarization makes Cyg X-1 challenging for detailed study of its bow shock structure.
The $q$ and $u$ maxima of Cyg X-1 occur at different orbital phases, which is expected from scattering of unpolarized radiation by an optically thin medium seen in Eq. (\ref{eq:QUfourier}).
This variability could correspond to a bow shock that reaches above the orbital plane with its lower half obscured by the accretion disk, but we refrain from drawing further conclusions from the data due to their low statistics.

Cyg X-3 may have more significant variability due to its higher wind mass-loss rate, which was estimated by \citet{Antokhin22} as  $\sim 10^{-5} ~ \mathrm{M}_\sun \mathrm{yr}^{-1}$, compared to $5 \times 10^{-7} ~ \mathrm{M}_\sun \mathrm{yr}^{-1}$ in Cyg X-1 \citep{Ramachandran2025}. 
The mass loss rate of Cyg X-1 was previously estimated by \citet{Gies03} to be an order of magnitude larger with the further expectation that the wind is directed towards the compact object \citep{Gies1986}.
Recently, \citet{Ramachandran2025} introduced a more comprehensive model with wind clumping and found no evidence for focused wind.
Clumping and wind asymmetry were not considered in \citet{Antokhin22}, so the mass loss rate of Cyg X-3 may also be overestimated.
It is accordingly difficult to estimate the density of the wind structures near either compact object, although the similarities in their orbital variability suggest that they might be roughly comparable.
Cyg X-3 is the only known HMXB with a WR donor in our Galaxy, so it may be the only easily observable source with pronounced orbital variability of X-ray polarization.
The opportunity to study the shock geometry in such detail is hitherto unique to Cyg X-3, making it a valuable tool in furthering our understanding of HMXB wind structures.

\section{Summary} \label{sect:summary}

We developed several analytical models for scattering in the intrabinary bow shock in Cyg X-3 and compared their predictions with the X-ray polarization data.
We find that the observed orbital variability can only be replicated if the bow shock is cylindrical, asymmetric across the shock apex, and at least somewhat optically thick.
The single-peaked orbital variability proved strongly constraining in determining the possible geometries.
Scenarios with the medium located directly above the obscuring funnel would dominate the scattered flux and lead to constant polarization, having difficulties with the high amplitude variations of polarization.
At the same time, any optically thin shock lying low on the orbital plane would lead to primarily two polarization peaks per orbit.
The observed coincidence of the maxima in q and u Stokes parameters can only be achieved, in the framework of these models, if the shock has asymmetry relative to its centerline.

The inferred shape of the bow shock confirms the shock geometry predicted by simulations of the HMXB jet-wind interaction.
The position angle of the orbital axis in our model fits deviates from the alignment of the innermost radio jet by a few degrees, which is consistent with an otherwise aligned jet with wind-induced bending.
As the nature of the variability does not change even when the jet is quenched, the funnel outflow must be key in forming the shock geometry.
Determining the exact properties of the shock and the impact of an inclined or curved jet would require more complex numerical modeling.

The high amplitude of the observed variability implies that the bow shock scattering is a significant fraction of the observed flux.
Previous models for the orbital light curve should therefore be revised with the addition of a variable component to the unabsorbed flux.
We have made a quantitative fit to the light curves and found that they can be explained by adding wind and clump absorption to the variable flux of our best-fitting model.
We thus expect the basic parameters of the system, such as its inclination, might change once the additional scattered flux is taken into account in the light curve modeling.


\begin{acknowledgements}
This work has been supported by a grant from the Turku University Foundation (VA).
AV acknowledges support from the Academy of Finland grant 355672.
Nordita is supported in part by NordForsk.
\end{acknowledgements}


\bibliographystyle{aa}
\bibliography{citations}

\begin{appendix}

\section{Plots of posterior distributions} \label{appendix:cornerplots}

\begin{figure}
\begin{center}
\includegraphics[width=\linewidth]{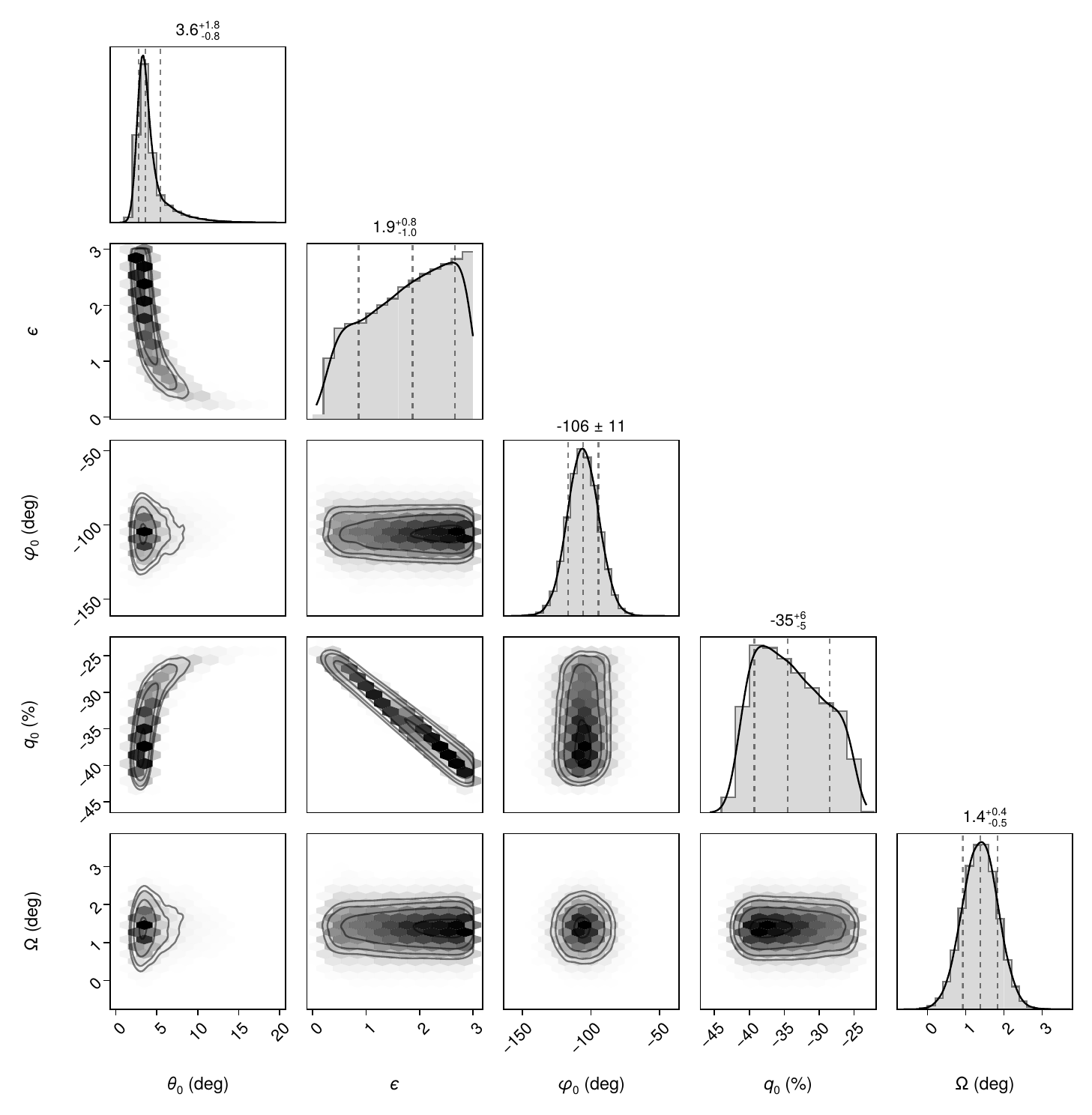}
\end{center}
\caption{Posterior distribution of the point-like cloud model fit.}
\label{fig:dotcorner}
\end{figure} 

\begin{figure}
\begin{center}
\includegraphics[width=\linewidth]{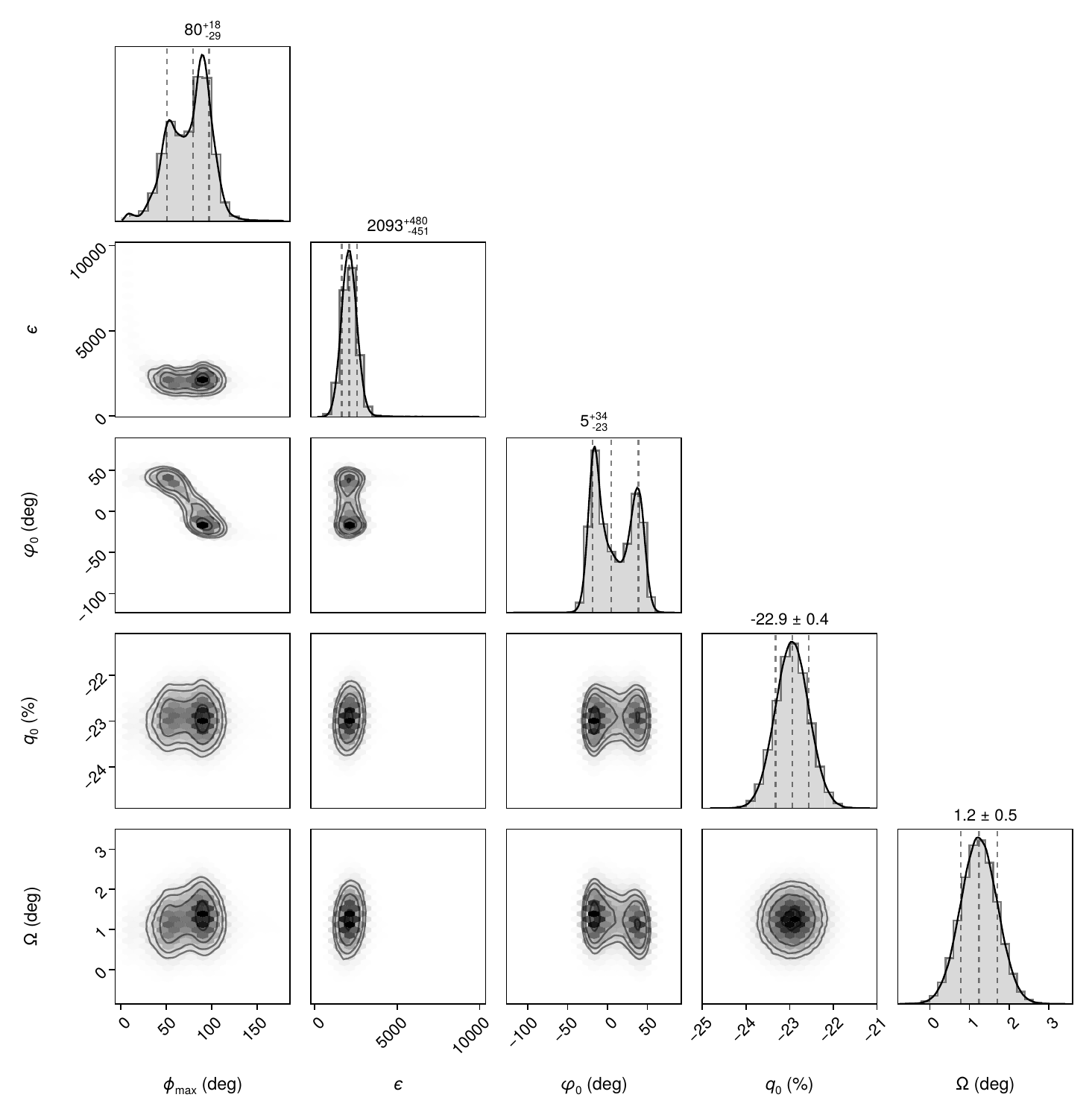}
\end{center}
\caption{Posterior distribution of the symmetric parabolic cylinder shock model fit.}
\label{fig:symcorner}
\end{figure} 

\begin{figure}
\begin{center}
\includegraphics[width=\linewidth]{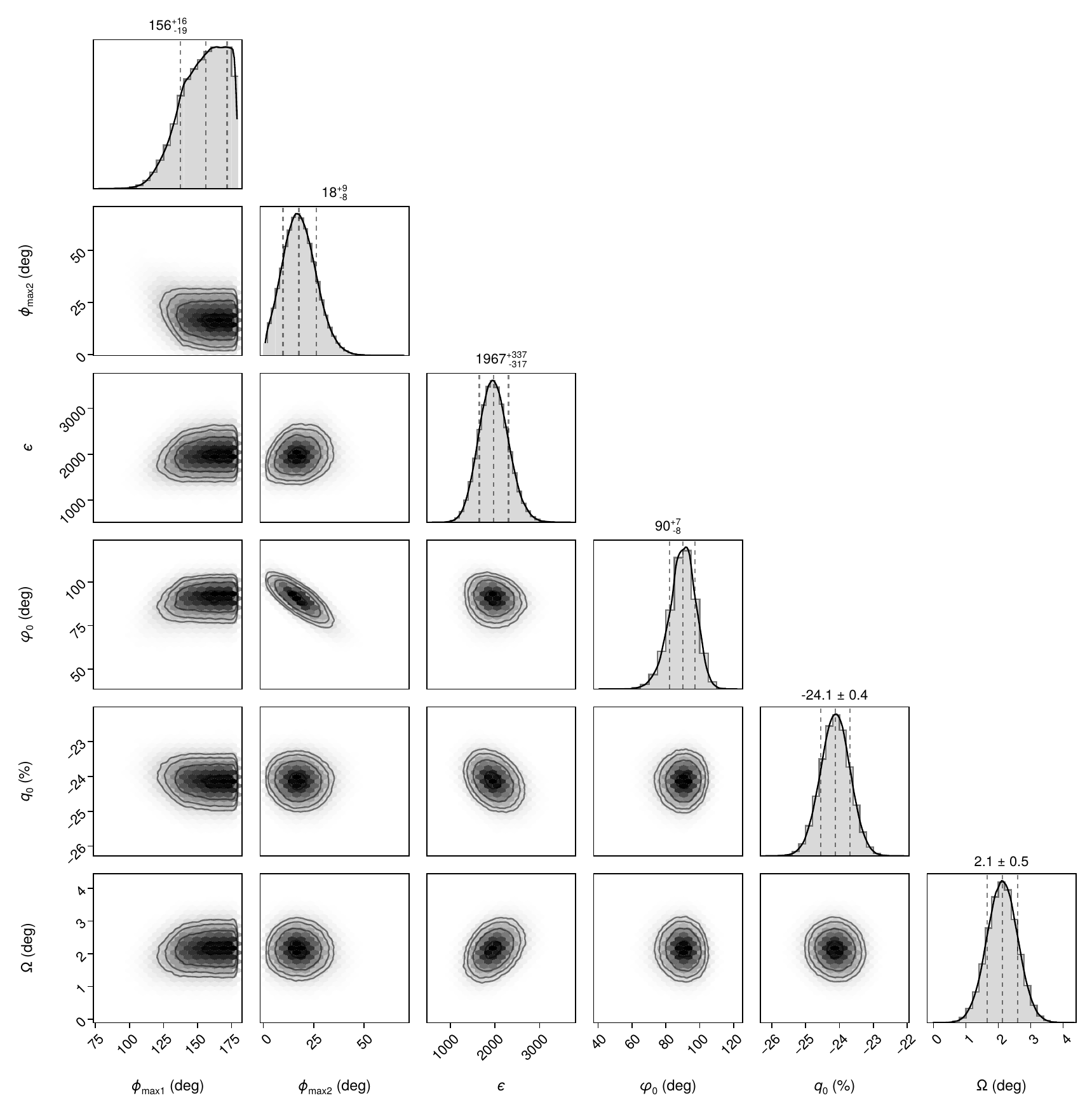}
\end{center}
\caption{Posterior distribution of the asymmetric parabolic cylinder shock model fit.}
\label{fig:asymcorner}
\end{figure}

\end{appendix}

\end{document}